# Eigenanalysis of morphological diversity in silicon random nanostructures formed via resist collapse

Makoto Naruse[1], Morihisa Hoga[2], Yasuyuki Ohyagi[2], Shumpei Nishio[2],

Naoya Tate[3], Naoki Yoshida[4] and Tsutomu Matsumoto[4,5]

[1] Network System Research Institute, National Institute of Information and Communications Technology, 4-2-1 Nukui-kita, Koganei, Tokyo 184-8795, Japan

[2] Dai Nippon Printing Co. Ltd., 250-1 Wakashiba, Kashiwa, Chiba 277-0871, Japan

[3] Graduate School of Information Science and Electrical Engineering, Kyushu University, 744 Motooka, Nishi-ku, Fukuoka 819-0395, Japan

[4] Graduate School of Environment and Information Sciences, Yokohama National University, 79-7 Tokiwadai, Hodogaya, Yokohama 240-8501, Japan

[5] Institute of Advanced Sciences, Yokohama National University, 79-5 Tokiwadai, Hodogaya, Yokohama 240-8501, Japan

Corresponding Author's E-mail: naruse@nict.go.jp

**Abstract:** This paper demonstrates eigenanalysis to quantitatively reveal the diversity and capacity of identities offered by the morphological diversity in silicon nanostructures formed via random collapse of resist. The analysis suggests that approximately $10^{115}$ possible identities are provided per 0.18-$\mu m^2$ area of nanostructures, indicating that nanoscale morphological signatures will be extremely useful for future information security applications where securing identities is critical. The eigenanalysis provides an intuitive physical picture and quantitative characterization of the diversity of structural fluctuations while unifying measurement stability concerns, which will be widely applicable to other materials, devices, and system architectures.

1. Introduction

Artifact metrics [1], also called physical unclonable functions (PUF) [2], utilize the unique physical properties of individual objects for authentication and clone resistance purposes in information security applications. These include electromagnetic [3], mechanical, and optical properties [4] associated with the objects such as ordinary paper [5], paper containing magnetic microfibers [6], plastics, and semiconductor chips. Matsumoto *et al.* demonstrates *nano-artifact metrics* that exploits physically uncontrollable processes occurring at the nanometer-scale, which are well beyond the scope of nanostructure technologies, in order to protect against increasing security attacks [7]. Versatile, nanoscale morphological patterns formed on the surfaces of planar silicon devices are one of the most unique and useful vehicles for nano-artifact metrics. This study investigates morphological diversity in silicon random nanostructures based on the eigenanalysis approach.

In particular, we utilize the random collapse of resists induced by exposure to electron-beam (e-beam) lithography [7]. Resist collapse may occur during the rinse process of lithography, and depends on the pattern resolution, resist thickness, and duration of e-beam exposure [8]. The result is the collapse of the intended pattern and production of versatile morphological patterns with minimum dimensions smaller than the resolutions of nanofabrication technologies [7]. The basic performance for security applications was clarified in Ref. [7] by quantitatively evaluating the false-match rate (FMR) for verifying identities, the false non-match rate (FNMR) for characterizing the stability of measurements, and the clone-match rate for evaluating the difficulties in making clones.

However, the technology's potential is not completely known yet, especially in terms of the number of different devices or identities that could be distinguished based on the nanoscale morphological patterns. This is especially important in view of future applications such as document security [9] and Internet-of-Things (IoT) [10], where securing identities of massive number of devices is critical. In this study, we demonstrate eigenanalysis of experimentally fabricated silicon random nanostructures, through which the diversity and the potential capacity of identities are quantitatively characterized. Our eigenspace-based approach provides intuitive physical pictures and quantitative discussions regarding the morphological diversity of nanostructured devices while uniting measurement stability, which is one of the important concerns for security applications.

**2. Experimental devices**

First, we review the experimental random nanostructured devices analyzed in this study. We fabricated an array of resist pillars with a 60 nm × 60 nm × 200 nm cross-sectional area on a grid of 120 nm × 120 nm squares that filled a 2 μm × 2 μm square, as shown in figure 1(a). We used an e-beam lithography system (JEOL JBX-9300FS) with a 100-kV acceleration voltage. In the rinse process, the random collapse of resist pillars was induced as shown in the scanning electron microscope (SEM) image in figure 1(b); versatile morphological patterns or structural fluctuations were observed. The resulting patterns were imaged by a critical-dimension SEM (CD-SEM, Hitachi High-Technologies CG4000). We analyzed 2383 samples fabricated on a single 200-mm–diameter wafer. Figures 1(c) and (d) show examples of SEM images of an array of collapsed resist pillars. The image contains 1024 × 1024 pixels and has an 8-bit resolution (256 levels). Versatile morphologies were observed, and the structural detail of the patterns (figure 1(c)) is as small as 9.23 nm [7].

3. Results and discussions

We examine experimentally observed structural diversity based on the following eigenspace-based principle. An 8-bit (256 levels) greyscale image of size 128 × 128 pixels was extracted from around the center of a pillar array image and smoothed using an 11 × 11 median filter. Here, one pixel occupies an area of approximately 3.3 nm side-length; therefore, the area of a single image is about 0.18 μm$^2$. We denote an image by a vector $\boldsymbol{x}_i$, which has $P = 128 \times 128$ elements. All the images or the total sets of devices are summarized by a matrix

$$Z = \{\boldsymbol{x}_1, \boldsymbol{x}_2, \cdots, \boldsymbol{x}_N\}, \tag{1}$$

where $N$ is the number of devices, which is 2383 in the present study. Let the mean values of all images be given by

$$\boldsymbol{\mu} = \frac{1}{N}\sum_{i=1}^{N}\boldsymbol{x}_i. \tag{2}$$

The covariance matrix is given by

$$C = \frac{1}{N}(Z - \boldsymbol{\mu})^T(Z - \boldsymbol{\mu}), \tag{3}$$

where $T$ indicates the matrix transpose and $(Z - \mu)$ implies subtracting the vector $\mu$ from each column of matrix $Z$. By solving the eigenequation

$$C\boldsymbol{s}_k = \lambda_k \boldsymbol{s}_k, \tag{4}$$

we derive the eigenvalues $\lambda_k$ and the eigenvectors $\boldsymbol{s}_k$. We set the index $k$ so that the eigenvalues are arranged in descending order. Such an eigendecomposition approach in analyzing a cluster of images is widely used in image processing literature [11] as well as in other high-dimensional systems such as robotic manipulators [12] and alignment of optical systems [13].

Figure 2(a) shows the cumulative sum of the eigenvalues divided by their sum. From this evaluation, 98% of the original images cannot be retrieved if we do not consider eigenvectors up to the 424th order, indicating that the morphological fluctuation of the devices under study spans a vast volume of high-dimensional eigenspaces (inset of figure 2(a)). Figure 2(b) shows the eigenvectors corresponding to the eigenvalues [from the 1st to the 10th], [from the 30th to the 39th], and [from the 400th to the 409th], each represented by a two-dimensional image. The eigenvectors with eigenvalues around the 400th order contain single-pixel-level resolution and fine-scale structures. From these observations, we can clearly grasp the physical versatility of silicon random structures by means of eigenanalysis. Furthermore, our interest is to quantitatively reveal the potential and the limitations of such morphological diversity.

Each image or device $\boldsymbol{x}_i$ is rewritten in the form of a linear combination of eigenvectors; $\boldsymbol{x}_i = \sum_k (\boldsymbol{x}_i \bullet \boldsymbol{s}_k)\boldsymbol{s}_k$. The projection of $\boldsymbol{x}_i$ to an eigenvector $\boldsymbol{s}_k$ is denoted by

$$X_k^{(i)} = \boldsymbol{x}_i \bullet \boldsymbol{s}_k, \tag{5}$$

meaning that $\boldsymbol{X}^{(i)} = (X_1^{(i)}, \cdots, X_P^{(i)})$ represents $\boldsymbol{x}_i$ in the eigenspace.

Here, an identical device can be observed in a slightly different manner depending on the observation conditions, which are affected by the misalignment or other reasons. Such fluctuations in measurements should be tolerated in order to ensure measurement stability, which is an important concern in view of security applications [7]. We quantify these fluctuations in measurements in the eigenspace as follows. Multiple measurements were performed for each of the 67 types of devices. Each measurement, for a certain identical device is denoted by $\boldsymbol{m}_i$ $(i = 1, \cdots, K)$, where $K$ is the number of measurements. The measurement $\boldsymbol{m}_i$ is represented in the eigenspace by

$$M_k^{(i)} = \boldsymbol{m}_i \bullet \boldsymbol{s}_k .\tag{6}$$

The measurement $\boldsymbol{m}_i$ is represented in the eigenspace by $\boldsymbol{M}^{(i)} = (M_1^{(i)}, \cdots, M_P^{(i)})$. Meanwhile, the $N$ types of individual device $\boldsymbol{x}_i$ $(i = 1, \cdots, N)$ are represented in the eigenspace by equation (5). The similarity between $\boldsymbol{M}^{(i)}$ and $\boldsymbol{X}^{(j)}$ is characterized by the inner product given by

$$R_{ij}^L = \sum_{k=1}^{L} M_k^{(i)} X_k^{(j)} ,\tag{7}$$

which should be maximum when $j$ is the designated identity that $\boldsymbol{M}^{(i)}$ belongs to. In equation (7), $L$ denotes the number of eigenvectors considered while evaluating similarity. If $L$ is too small, similarity does not provide an accurate classification; we quantify the minimum dimensions in identifying individual devices without errors. The ratio of inadequate classification or equivalent FNMR is evaluated for all 67 devices as a function of the dimension ($L$). Figure 3(a) shows six representative cases; the circular marks indicate that $L$ should be larger than 15 to suppress the error rate to zero. Concerning all devices, the error rate results in zero, if $L$ is larger than 38.

Furthermore, we consider the system's capacity. Let a measurement $m_i$ $(i = 1, \cdots, K)$ or equivalently its eigenspace representation $M^{(i)}$ belong to the identity $j$ represented by $x_j$ or $X^{(j)}$ in the eigenspace. The deviation of $m_i$ around the identity $x_j$ is characterized in the eigenspace by

$$M^{(i)} - \left( \sum_{k=1}^{L} M_k^{(i)} \hat{X}_k^{(j)} \right) \hat{X}^{(j)}, \tag{8}$$

which represents the deviation of $M^{(i)}$ from the projected vector of $M^{(i)}$ on to $\hat{X}^{(j)}$, as schematically shown in figure 3(b). Here, $\hat{X}^{(j)}$ is the normalized vector of $X^{(j)}$. By calculating the deviation vector given by equation (8) for all measurements $i = 1, \cdots, K$, we derive a vector representing the mean value of all measurements, denoted by $S = (s_1, s_2, \cdots, s_P)$, which characterizes the extent to which individual devices should span in the eigenspace while tolerating the measurement stability. Considering that the first $L$ (= 38) eigenvectors are necessary to satisfy the measurement stability, the volume spanned by $S$ is characterized by the multiplication of the first $L$ elements $\prod_{k=1}^{L} s_k = s_1 \times s_2 \times \cdots \times s_L$ [12]. Taking the ensemble average over 67 devices, $\left\langle \prod_{k=1}^{L} s_k \right\rangle$ is estimated to about $8.11 \times 10^{75}$.

Meanwhile, the volume of the eigenspace spanned by different types of devices is characterized by the multiplication of the eigenvalues $\prod_{k=1}^{L} \lambda_k = \lambda_1 \times \lambda_2 \times \cdots \times \lambda_L$ [12], which is approximately $7.29 \times 10^{190}$ in the present study. Such a metric is also known to be useful in characterizing the manipulability of robots [12]. By dividing this amount by the individual device's volume, as schematically shown in figure 3(c), the capacity or the number of potential identities is derived by $\left.\prod_{k=1}^{L}\lambda_k \middle/ \left\langle \prod_{k=1}^{L} s_k \right\rangle\right.$, which is estimated at approximately $10^{115}$ or equivalently approximately 380 bits. This amount is significantly large, achieved by a 0.18-$\mu m^2$ device indicating the huge capacity of the morphological patterns of silicon nanostructures formed via resist collapse. In other words, the eigenspace-based insight into structural diversity observed in silicon structures while unifying measurement tolerances provides an intuitive physical picture as well as quantitative estimations of their potential performances.

**4. Conclusion**

In summary, we demonstrate the eigenanalysis of morphological fluctuations in silicon nanostructures formed via resist collapse in order to clarify its potential for nano-artifact metrics or PUF. The result shows that the nanoscale morphological patterns are widely distributed in a hugely high-dimensional eigenspace; 424th order eigenvectors are needed to reconstruct 98% of the original nanostructured patterns in our particular experiment. Considering measurement stability in the eigenspace, the estimated number of implementable different identities is approximately $10^{115}$ in a 0.18-$\mu m^2$ nanostructure area, indicating the usefulness of nanoscale devices to accommodate huge amounts of uniqueness for future security applications. The presented approach provides an intuitive physical picture regarding the diversity of structural fluctuations and its systematic and quantitative characterizations, which will be widely applicable to other materials, devices, and system architectures.

**Acknowledgments**

This work was supported in part by the Grant-in-aid in Scientific Research and the Core-to-Core Program, A. Advanced Research Networks from the Japan Society for the Promotion of Science.

**Figure captions**

**Figure 1.** (a,b) The design of a resist pillars array, (a) which is collapsed randomly during the rinse process of e-beam lithography as observed in a SEM image shown in (b). (c,d) SEM images of the fabricated devices (total number: 2383) exhibiting versatile morphologies due to the random collapse. The minimum dimension is occasionally <10 nm, (c) which is considered extremely difficult to fabricate clones. Such versatilities, derived by physical randomness occurring in dimensions beyond the expected technology roadmap, are the fundamental ideas of nano-artifact metrics or PUF [7] (adapted by permission from Macmillan Publishers Ltd: Scientific Reports [7], Copyright 2014).

**Figure 2.** Eigenanalysis of the fabricated devices. (a) Cumulative sum of the eigenvalues divided by their sum as a function of the eigenvalues' number. In retrieving 98% of the individual images, 424 eigenvectors are required, indicating that the fabricated morphological patterns span in a high dimensions in the eigenspace. (b) Examples of two-dimensional

representations of the eigenvectors corresponding to the eigenvalues [from the 1st to the 10th], [from the 30th to the 39th], and [from the 400th to the 409th].

**Figure 3.** (a) Eigenanalysis of the measurement stability: Multiple measurements for an identical device should yield the same device identification. When the number of eigenvectors considered during identification decreases, the identification error, referred to as the false no-match rate (FNMR), increases. When more than 38 eigenvectors are considered, FNMR gives zero for all device sets. (b) System capacity estimation: considering the deviation of each measurement from the specified identity, the total amount of deviations for multiple measurements is evaluated in the eigenspace. (c) The potential number of identities is estimated by dividing the volume of the total eigenspace by the volume of the individual devices to tolerate the measurement stability.

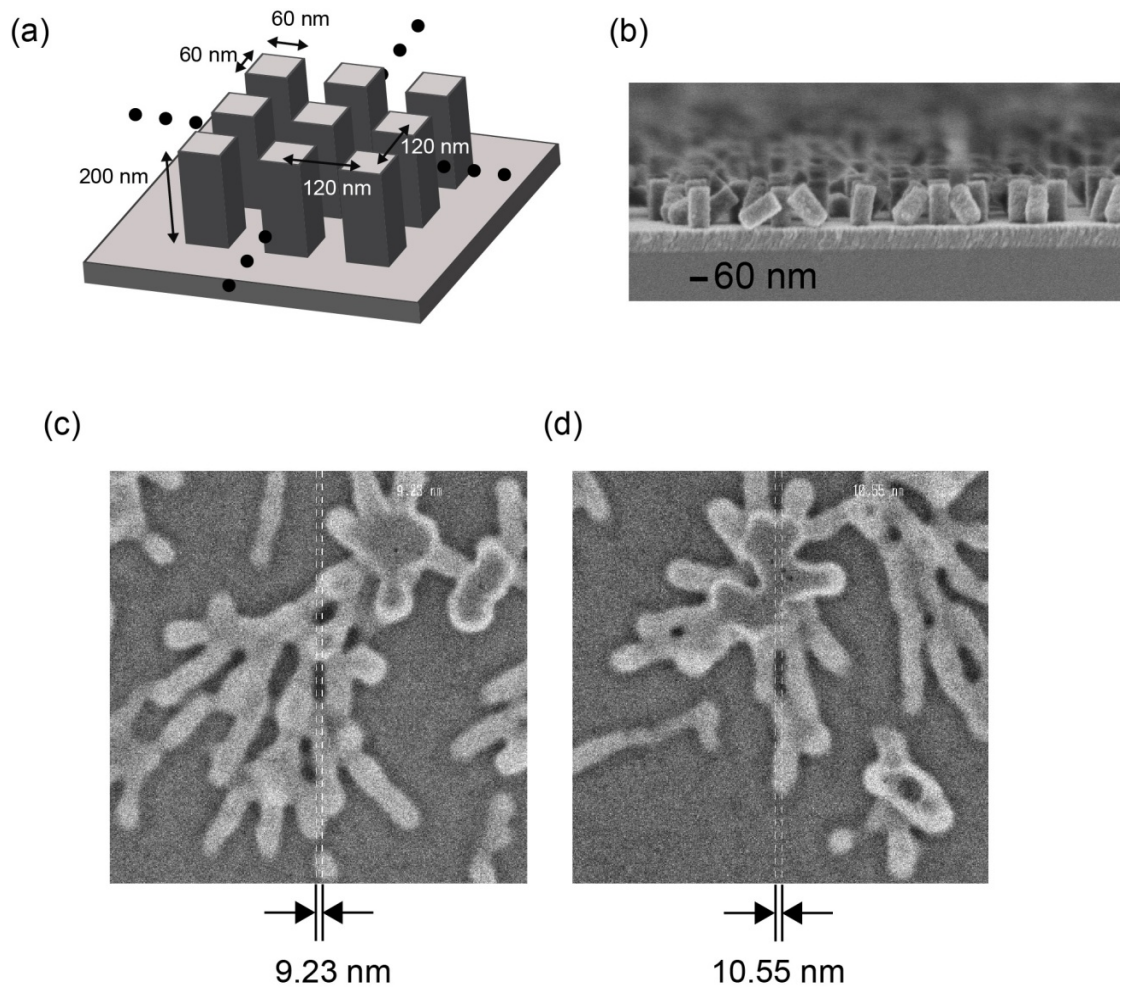

Figure 1

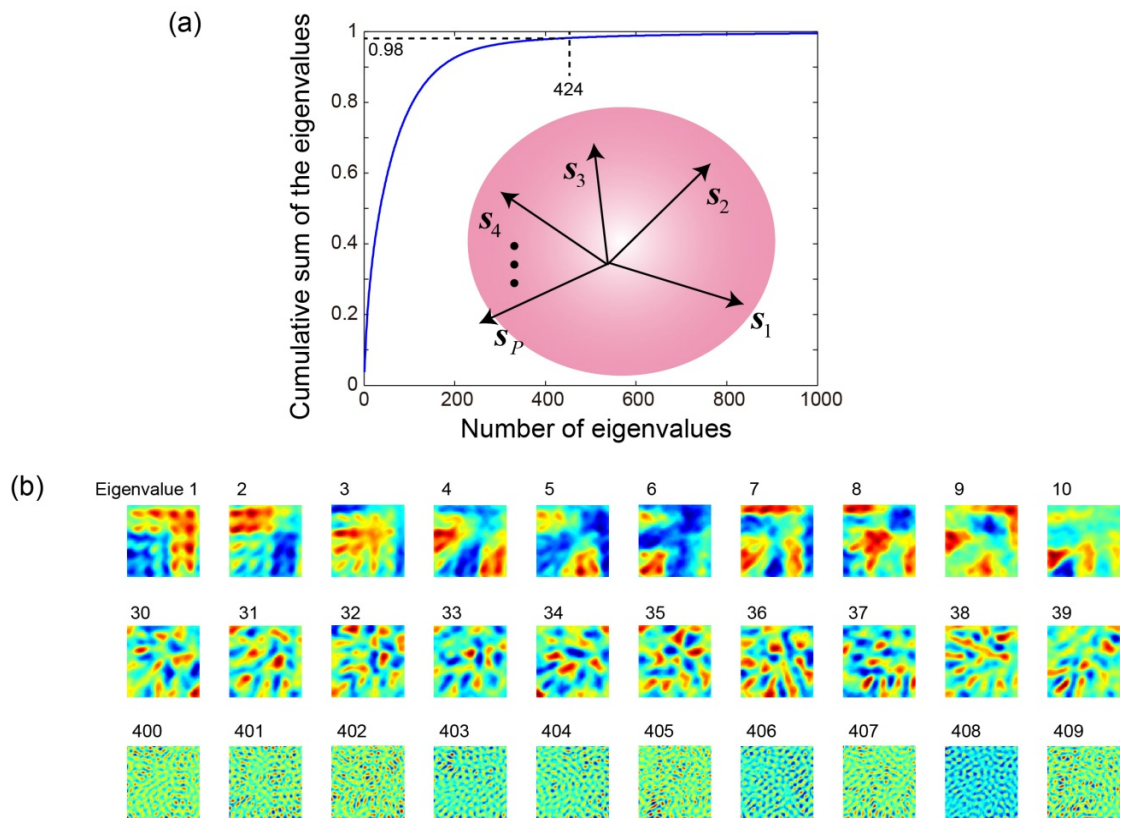

Figure 2

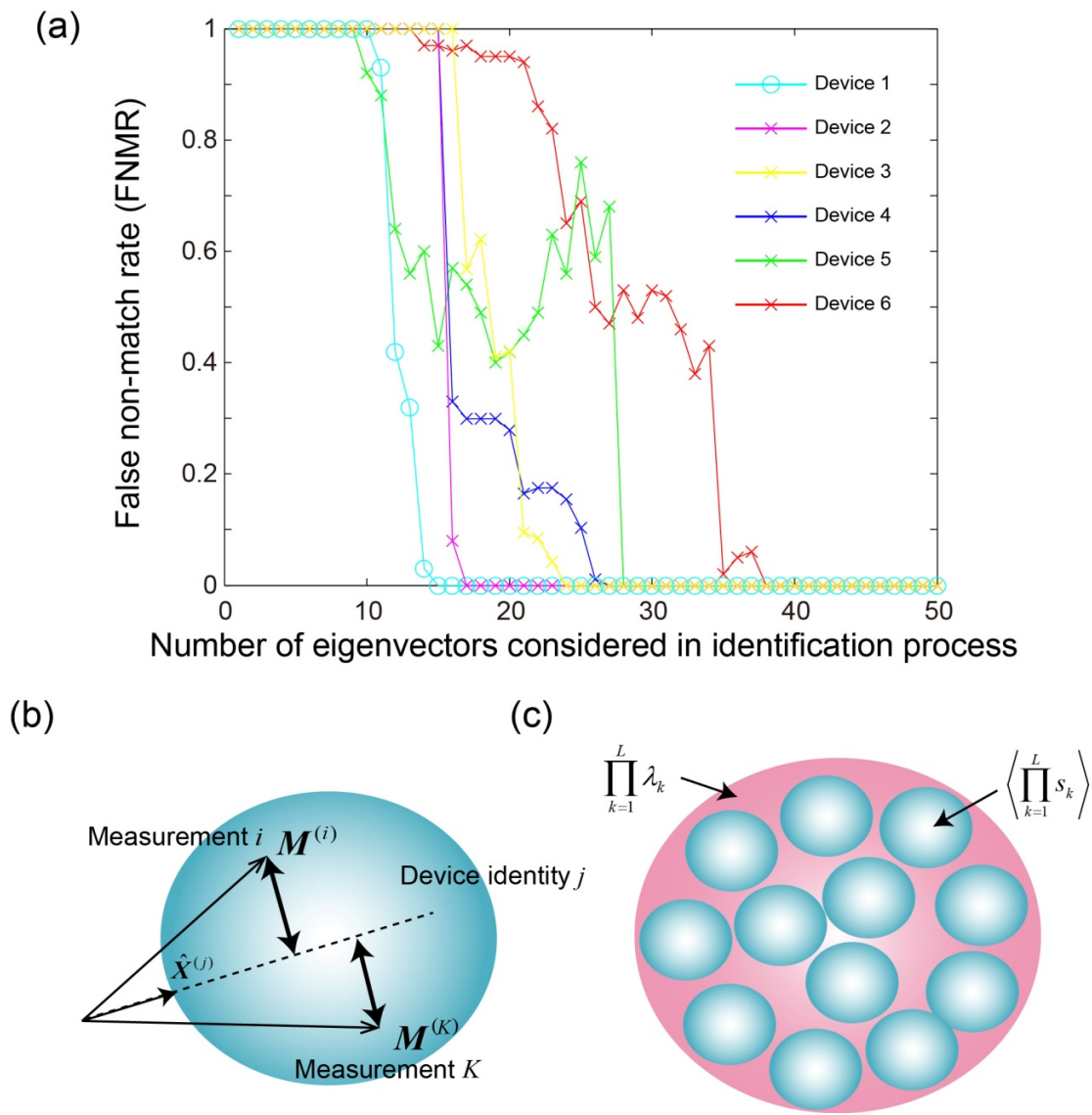

Figure 3